\documentclass[12pt,preprint]{aastex}
\begin{document}
\title{Determinations of temperature and density for solar-like Stars
using Si XI soft X-ray emission lines}
\author{G. Y. Liang, G. Zhao}
\affil{National Astronomical Observatories, Chinese Academy of
Sciences, \\ Beijing 100012, P. R. China} \email{gzhao@bao.ac.cn}

\begin{abstract} We study temperature and density sensitivities of
ratios of Si XI soft X-ray emission lines, in the wavelength range
of 43---54\AA\,. The typical temperature of the formation of the
analyzed lines is around 1.6~MK, which makes this analysis
complementary to the analysis of He-like triplets being sensitive
to hotter plasma. We present theoretical calculations and compare
them with ratios obtained from high-resolution X-ray spectra of
five solar-like stars: Procyon, $\alpha$~Cen~A$\&$B,
$\epsilon$~Eri, and Capella. We find that our results are in good
agreement with results obtained by other authors through different
diagnostics, namely the analysis of density- and
temperature-sensitive He-like triplet lines. We further estimate
the coronal pressure and filling factors from Si~XI lines in this
study.
\end{abstract}


\section{Introduction}
Determinations of elemental abundances and structures in a various
of stars belong to the most important tasks of observational
astronomy and the understanding of the evolution of chemical
elements. The temperature structure of emitting layers such as the
stellar coronae must be known before elemental abundances can be
determined. The density is another important physical parameter to
describe the magnetically confined plasma in outer atmospheres of
late-type stars, which can be used to estimate the spatial
information for the stellar coronae. For those pre-main sequence
stars, the constrained density can give us clues for the origin of
the X-ray production such as magnetics alike the stellar coronae
or a shock due to accretion processes.

Before the launch of missions with high resolution spectrograph
such as the $Chandra$ and XMM-$Newton$, a determination of the
electron temperature of X-ray emitting layers usually adopts a
technique of global-fitting to spectra. The determined temperature
structure from this technique strongly depends on the
determination of continuum level of the observed spectra. The
measurement of electron density is not possible for non-solar
stars with previous X-ray satellites such as ASCA and ROSAT
configured with low- and medium-resolution spectrographs. The
observation data provided by high-resolution spectrograph on board
new generation satellites $Chandra$ and XMM-$Newton$, allow us to
estimate this parameter. Individual emission lines can be resolved
from high-quality spectra, further the electron temperature can be
derived from line intensity ratios such as $G$ value (between
intercombination $i$ and forbidden $f$ lines {\it vs} resonance
line $r$) of He-like ions which was described by \citet{PMD01} in
detail. Since then, the estimation of the electron density becomes
possible from X-ray emission line ratios such as $R$ value between
$i$ and $f$ lines He-like ions. Using observation data with lower
spectral resolution, $n_{\rm e}$ and emission measures $EM$ can
not be obtained independently, such that no emitting volumes $V$
could be estimated from the relation of $EM=n_{\rm e}^2~V$. The
structural information of the outmost atmosphere of stars was
accessible from the analysis of lightcurves for a few special
systems such as rotational modulation, eclipse mapping, etc.
However, the solution of the reconstructed differential EM is
always nonunique, and involves the difficulty of disentangling
time-dependent versus geometric variability
\citep{SDH96,SF99,Gue04}. Since the operation of the two
satellites with high-resolution and high-collecting areas, the two
parameters have been derived from spectroscopic methods for every
class of X-ray sources by many observers
\citep[etc.]{BGK00,CHD00,KHS02,NSB02,NBD03,RNM03,SS04,SRN05}.
Using He-like triplet lines, a systematic investigation of the
electron temperature and density for late-type stars with various
levels of activity, has been made by \citet{NSB02,Ness04} and
\citet{Testa04}. However, the UV field significantly affects the
He-like triplet lines. Recently, \citet{NS05} presented a density
diagnostic ratio being independent from the UV field, that
therefore allows to disentangle $n_{\rm e}$ and UV field that both
affect the O~VII triplet lines. Additionally, the derived density
has also been used to estimate the production mechanism (alike
stellar corona or accretion) of the X-ray emission in pre-main
sequence stars \citep{SS04,SRN05}.

Soft X-ray emissions of L-shell ions of silicon have been detected
extensively in spectra of stellar coronae. These emissions carried
out information of the two physical parameters for the solar-like
coronae. In our previous work, we investigated the emission lines
of Si X, and diagnosed the coronal electron density using them for
late-type star---Procyon \citep{LZS06a,LZS06b}. The soft X-ray
emissions arising from 3d---2p, 3s---2p and 3p---2s transitions of
Si XI have also been detected obviously for some cool stars, such
as Procyon and $\alpha$ Cen A\&B \citep{RMA02,RNM03}. And the
temperature (1.6~MK) of maximum formation of Si~XI ion in the
ionization equilibrium \citep{MMC98}, is the typical temperature
of corona of stars with low activities, which makes the analysis
of Si~XI soft X-ray emission lines interesting.

EUV features arising from 2s$^2$--2s2p and 2s$^2$--2p$^2$
transitions of Si XI have been used to derive the electron
temperature and density for solar flares. \citet{KGF95} presented
seven line ratios whereas these ratios are usually sensitive to
both the electron temperature and density of the emitting plasma.
Moreover measured ratios obtained with the Naval Research
Laboratory's S082A spectrograph on board {\it Skylab}, are larger
than theoretical high-temperature and high-density limits
\citep{KGF95}. Later, \citet{LBM01} calculated these line ratios
again and compared with solar observations obtained with normal
incidence spectrometer on board SOHO CDS, as well as laboratory
measurements performed by \citet{KKK96}. In contrast to the EUV
lines, there has been little work on the soft X-ray transitions of
Si XI in astrophysical literatures, probably as a result of the
limited availability of high quality spectra of stellar in this
wavelength region.

This paper is organized as following, we firstly investigate the
Si~XI soft X-ray emission lines and present theoretical
calculations of temperature- and/or density-sensitive line
intensity ratios in Sect. 2. A brief description of observations
and extractions of line fluxes is presented in Sect. 3. Results
and discussions are outlined in Sect. 4. Our conclusions are given
in Sect. 5.

\section{Theoretical line intensity ratios}
Our model of Si XI consists of energetically lowest 350
fine-structure energy levels belonging to configurations 2s$^2$,
2s2p, 2p$^2$, 2l3l$'$, 2l4l$'$, 2l5l$'$, 2l6l$'$ and 2l7l$'$ (l=s,
p; l$'=0, 1, ..., n-1$). Some energy levels are replaced by
available experimental values from the NIST
database\footnote{http://physics.nist.gov/cgi-bin/AtData/main$\_$asd}
and results of \citet{CT01}. The theoretical calculation is
performed using a fully relativistic method---flexible atomic code
(FAC) developed by \citet{Gu03,Gu04}. Self-consistent results of
electron impact excitation, de-excitation rates and Einstein
$A$-coefficient (including electric- and magnetic-dipole, as well
as quadrupole transitions) among the 350 levels were also included
in present model. The input atomic data of our model significantly
differs from that of CHIANTI code, in which only 46 low-lying
energy levels, and the excitation and radiative decay among the 46
levels were included. Present model is an extensive one, and
cascade effects on the upper levels of interested transition lines
from higher excited levels up to $n=7$ ($n$ is the main quantum
number), have been considered. In our recent work \citep{LZZ06c},
a detailed assessment for the calculated data is performed, which
reveals that the atomic data is reliable for diagnostic
application. Here, a brief comparison of our data with the input
atomic data of CHIANTI code is made as shown in Fig. 1. The top
panel of Fig. 1 illustrates the comparison of different
theoretical predictions of energy levels with the available
experimental ones (the NIST data and that of \citet{CT01}). This
panel indicates that our energy levels agree with the experimental
values within 0.7\%, which is better than the theoretical data in
the CHIANTI database. NIST only lists the energy levels being less
20~$Ryd$. For higher energy levels of above $\sim$20{\it Ryd},
\citet{CT01} reported some values and determined the energy levels
from observed wavelengths by an interactive optimization procedure
using the program ELCALC \citep{RK69}. Comparing with their energy
levels also can benchmark our prediction in a certain extent. A
better agreement is shown by the filled circles in the upper panel
of Fig. 1. The comparison of weighted oscillator strengths is
illustrated in the middle panel of Fig. 1, which reveals the two
different data is comparable within 20\% for most strong
transitions ({\it gf}$>$0.1). For those transitions involved in
below interested ratios (filled up-triangle in Fig. 1), the two
different predictions show a good agreement within 20\%. The
prominent 3d--2p transitions exhibit a better agreement. For some
weak transitions ({\it gf}~$<$0.1), large discrepancies up to
several factors appear. We suggest such differences are due to
different inclusions of configuration interaction (CI) in the
different data sources. The work of \citet{AKN05} and
\citet{ZLG05} has revealed that the consideration of an elaborate
CI is very necessary, which support our performance of atomic data
calculation. For the effective collision strength ($\Upsilon$)
obtained under the Maxwellian distribution, only the data at a
temperature of 1.0~MK (the bottom panel of Fig. 1) is compared for
concision, because the temperature is the typical coronal
temperature of stars with less activity.

Proton impact excitation is important process for population and
de-population of the lowest levels with $n=2$ complexes. Data from
\citet{RFC98} was employed in present model, in which the
close-coupled impact-parameter method was adopted in the
calculation.

The theoretical line intensities are obtained by solving
steady-state rate equations with optically thin assumptions, as
described by \citet{Gu03} and \citet{BCK01}. Such assumptions are
based on the following reasons. Firstly, photon-excitation and
de-excitation rates are negligible in comparison with the electron
impact excitation rates in stellar corona, tokamak and
laser-produced plasmas, so the processes aren't considered;
secondly, the rates of ionization to and recombination from other
ionic stages are lower than bound-bound transition rates; finally,
emitted photon directly escapes the plasma, does not collide with
particles again. The line intensities are calculated over a wide
range of the electron temperature and density. Here, we pay
careful attention on five prominent lines of Si XI, as listed in
Table 1. The $3s-2p$ line at 52.306\AA\, is a strong and isolated
line in these selected features. A detailed analysis indicates
that following emission line intensity ratios of Si XI ion are
sensitive to the electron temperature and/or density:
\begin{eqnarray}
R_1 &= &I(52.306~{\rm \AA})/I(43.743~{\rm \AA}), \nonumber \\
R_2 &= &I(52.306~{\rm \AA})/I(49.207~{\rm \AA}), \nonumber \\
R_3 &= & I(49.207~{\rm \AA})/I(46.391{\rm \AA}), \nonumber \\
R_4 &= & I(52.306~{\rm \AA})/I(46.391~{\rm \AA}), \nonumber \\
R_5 &= & I(43.743~{\rm \AA})/I(46.391~{\rm \AA}), \nonumber \\
R_6 &= & I(46.391~{\rm \AA})/I(46.283~{\rm \AA}), \nonumber
\end{eqnarray}
and
\begin{eqnarray}
R_7 &= & I(52.306~{\rm \AA})/I(46.283~{\rm \AA}). \nonumber
\end{eqnarray}
Therefore they may be useful diagnostic techniques for the
electron temperature and density for all kinds of hot plasma.

Fig. 2 plots these ratios as a function of the electron
temperature at $n_{\rm e}$=1.0$\times$10$^9$~cm$^{-3}$, and {\it
vs} the electron density at log$T_{\rm e}$(K)=6.2 which is the
temperature of maximum formation of Si~XI(see solid line in the
figure). The ratios are given in photon units, and used throughout
the paper. The ratio $R_1$ varies from 1.7 to 0.9 over the
temperature range of log$T_{\rm e}$(K)=6.0---6.4, and no variation
occurs through a large range of density
10$^7$---10$^{12}$~cm$^{-3}$, which is the typical condition of
astrophysical, tokamak and EBIT plasmas. This indicates the ratio
$R_1$ is a powerful diagnostic technique of the electron
temperature for the hot astrophysical and laboratory plasmas.
Ratio $R_2$ exhibits the same behavior as $R_1$, yet this ratio
shows a less temperature-sensitivity than that of $R_1$.

The ratio $R_3$ appears to be sensitive to the electron density
(shown in Fig. 3), simultaneously it displays a certain
sensitivity to the electron temperature. For example, increasing
log$T_{\rm e}$(K) from 6.2 to 6.4 leads to 8~per~cent variation at
$n_{\rm e}$=1.0$\times$10$^9$~cm$^{-3}$. The feature at
46.391~\AA\, being correlative to ratios $R_3$, $R_4$ and $R_5$
has been identified to be 46.410~\AA\, line arising from ${\rm
2s2p~^3P_2 \to 2s3d~^3D_3}$ transition of Si XI in work of
\citet{RMA02}. After adjusting by experimental value from the NIST
database, a better wavelength 46.401~\AA\, is obtained in present
work. We noticed that the emissivity of the line at 46.404~\AA\,
is about 10~per~cent relative to that of 46.401~\AA\,, whereas the
two lines can not be resolved using present available
spectrographs. The expected blending of 46.404~\AA\, line is
rather smaller than both the theoretical uncertainty and
statistical error in the line flux measurement, however the
blending effect was considered in this study. Ratio $R_4$ is also
sensitive to the electron density, but it is insensitive to the
electron temperature in the density-sensitive region, as shown in
the lower panel of Fig. 2. The variation is less than 4~per~cent
when log$T_{\rm e}$(K) changes 0.2~dex. This indicates $R_4$ is a
powerful $n_{\rm e}$-diagnostic tool for the line formation
region. As $R_3$, the blending effect in 46.391~\AA\, line has
been also considered in the prediction of $R_4$. Ratios
$R_5$---$R_7$ show obvious $n_{\rm e}$ and $T_{\rm e}$-dependence,
as shown by curves of $R_5$ and $R_6$ in Fig. 2. For clearness,
the curve of $R_7$ {\it vs} $T_{\rm e}$ is not given, only its
dependence on $n_{\rm e}$ is plotted in Fig. 3. The resolved lines
and their blending contributions are listed in Table 1. From above
analysis, we distinguish that $R_1$ and $R_4$ are powerful $T_{\rm
e}$ and $n_{\rm e}$ diagnostic method, respectively. Using the
properties of Si~XI, we can estimate the electron temperature and
density in Si XI emitting region for cool stars.

\section{Observations and line fluxes}
In this work, we re-analyze the spectra of stars including three
normal dwarf stars, i.e., $\alpha$~Cen~A\&B, Procyon and
$\epsilon$~Eri, and an active binary system Capella. The detailed
descriptions and analyses for these stars are available from many
literatures \citep[etc.]{SDH96,MRD01,ABG01,NMS01,RMA02,RNM03}.
Here, we briefly describe the observations and data reductions for
these stars. $\alpha$~Cen~A and B are firstly resolved from
spectral observation with Low Energy Transmission Grating
Spectrometer (LETGS) instrument on board {\it Chandra}
observatory, a detailed description about it was given by
\citet{RNM03}. All observations use the LETGS combined with High
Resolution Camera (HRC) detector. Reduction of the LETGS datasets
adopts CIAO3.2 software with the science threads for LETGS/HRC-S
observations. For the extraction of the spectra of
$\alpha$~Cen~A\&B, a similar procedure used by \citet{RNM03} is
adopted here under the CIAO environment.

Line fluxes are obtained by locally fitting the emission features
using multiple overlapping Gaussian profiles, together with a
constant value representing background and (pseudo-)continuum
emission, which was determined in line-free regions such as
(51---52\AA\,). The positive order spectra was used to derive the
line fluxes because some lines such as 52.306~\AA\, of Si XI lie
in the gap between chips in the negative spectra. Determinations
of the line fluxes of Si XI lines are difficult because a large
amount of emissions of highly ionized magnesium, sulfur and argon
is also present in this wavelength range of 43---53\AA\,. Here we
choose Procyon as an example to illustrate. At the shorter
wavelength wing of Si~XI line at 43.743~\AA\,, a weak line of
S~XII at 43.649~\AA\, appears as illustrated in the left panel of
Fig. 4. Two emission lines of S~IX at 49.107~\AA\, and
49.324~\AA\, broaden the shoulder of Si~XI line at 49.207~\AA\, as
shown in the right panel of Fig. 4. By the best-fitting to the
spectra, observed line fluxes are obtained as listed in Table 1
with 1$\sigma$ statistical error. In this table, labels such as
$3a$ and $3b$ in first column denote this line is blended. For
Procyon and $\alpha$~Cen~A\&B, all prominent lines of Si XI have
been identified by \citet{RMA02,RNM03}, and line fluxes are
consistent with those derived by us within 1$\sigma$ statistical
error. However, for $\epsilon$~Eri and Capella, only the line at
49.207~\AA\, is reported in available literatures.

\section{Results and Discussions}
\subsection{Determinations of $T_{\rm e}$}
Sect. 2 has revealed that $R_1$ and $R_2$ are good $T_{\rm e}$
diagnostic tools. By comparing observed line ratios $R_1$ and
$R_2$ with the theoretical predictions as shown in Fig. 5, the
electron temperature of the Si XI emitting region is obtained, as
listed in Table 2. The observed ratio $R_1$ of $\alpha$~Cen~A
slightly higher than other observed ratios, and the value is up to
1.72. This implies that the diagnosed temperature is lower than
that of other stars. In other cases, the observed ratios of $R_1$
are ranging from 1.2 to 1.4. In consequence, the diagnosed
electron temperatures lie within $\sim$1.5--2.4~MK. These results
are consistent with the temperature (1.6~MK) of maximum formation
of Si XI in the ionization equilibrium \citep{MMC98}. The results
also agree with the peak temperature of differential emission
measure (DEM) constructed from a global spectral fitting for
normal stars including Procyon and $\alpha$~Cen
\citep{RMA02,RNM03}, and low-temperature component of DEM for
active binary system---Capella \citep{ABG01}. For $\alpha$~Cen
binary system, our results further confirms Raassen's conclusion
derived from He-like triplet that $\alpha$~Cen~B (K1V) component
is hotter than the $\alpha$~Cen~A (G2V) component \citep{RNM03}.

As stated in Sect.2, $R_2$ shows a less sensitivity than that of
$R_1$, so measured ratios are compatible with the upper limit of
the theoretical prediction within 1$\sigma$ statistical error. In
the case of Procyon, \citet{RMA02} have reported that the Si~XI
line at 49.207\AA\, is contaminated by Ar IX line at 49.181~\AA\,
which can not be resolved out by present LETGS instrument on board
{\it Chandra} observatory. Therefore, the correct extraction of
the fraction of Ar~IX line is very important. Fortunately,
\citet{LBB03} also detected this line in their recent experiment
conducted in EBIT measurement for L-shell spectra of highly
charged argon. Using the experimental ratio (1.11) between the
lines at 48.737~\AA\, and 49.181~\AA\, of Ar IX measured in the
EBIT facility, we extracted the contribution of Ar IX line and
rederived the ratio $R_2$. The weak $T_{\rm e}$-dependence of the
ratio of the two Ar~IX lines supports the direct application of
the experimental value. Taking Procyon star as an example, the Ar
IX line at 48.730~\AA\, has been obviously detected and no other
contamination. We conclude that the Ar IX line at 49.181~\AA\,
occupies about 18~per~cent to the total flux around 49.207\AA\,.
The intensity of the Ar~IX line is nearly 2$\sigma$ errors. By
considering the $\sim$18~per~cent contribution, the ratio $R_2$
increases from 0.52$\pm$0.13 to 0.64$\pm0.14$, accordingly the
derived temperature drops from 7.07 to 2.24~MK which shows
consistency with that derived from $R_1$ within error. For our
other sample, we measured the line flux of Ar~IX around
48.730\AA\,, and performed similar procedure as in Procyon. The
derived temperatures show an good agreement with those from $R_1$
within $1\sigma$ error. We also notice that the intensity of Ar~IX
line relative to Si~XI lines changes with the coronal temperature,
and specifically, a lower contamination from Ar~IX lines is
expected for more active stars.

\subsection{Determinations of $n_{\rm e}$}
Sect. 2 indicated that $R_4$ is a powerful diagnostic tool for the
electron density in the Si XI emitting region of hot plasma. By
comparing the observed ratios with theoretical prediction, we
derived the electron density in the Si XI emitting region of
coronal plasma, as shown in Fig. 5. Here, the comparison of $R_4$
together with $R_3$ is illustrated in Fig. 6. Table 3 lists the
observed line ratios with 1$\sigma$ statistical error and the
diagnosed density for our sample.

The first impression of Fig. 6 is the uncertainty of observed line
ratios appearing to be large, and almost all the observed ratios
of $R_4$ agreeing with high-density limit in statistical error
with exception of $\alpha$~Cen~B case being agreement with
low-density limit. A long exposure times are necessary to better
constrain the density from $R_4$. Yet present line ratio
measurements still can give us some clues for the Si XI emitting
region, that is the lower or upper density limit can be
constrained from our measurements. We note that mean value of the
electron density for our sample is above 2$\times10^8$~cm$^{-3}$,
which is the typical density of solar quiescent coronal plasma.
For the ratio $R_3$, the contaminated Si~XI line at 49.207\AA\, is
concerted, so the blending effect due to the Ar~IX line is
considered as above subsection. Taking Procyon as an example,
$R_3$ drops from 3.60$\pm$1.07 to 2.95$\pm0.86$ by taking into
account the contamination. A density of
$<8.8\times10^{10}$~cm$^{-3}$ is obtained, which agrees with that
constrained by $R_4$ within $1\sigma$ statistical error. Since the
large statistical errors, only the lower or upper limit of the
density is given for our sample as listed in Table 3.

\subsection{Comparison between present and published results}
The temperature of maximum formation of Si XI is close to that of
N VII emission lines, therefore the physical conditions in Si XI
and N VII emitting region can be compatible. For N VII, the
temperature can be derived in principle from ratio between
Ly$\alpha$ line and Ly$\beta$ line. However the Ly$\beta$ line
generally is faint. So the temperature derived from ratio between
Ly$\alpha$ line and resonance line of He-like ion can trace the
this information. Firstly, we compare our results of $T_{\rm e}$
with those derived from emissions of H- and He-like N ions as
shown in Table 4. In this table, Ly$\alpha/r$ column represents
temperature derived from ratio between Ly$\alpha$ and resonance
line $r$ of He-like N VI, while the next column represents
temperature constrained by ratio between Ly$\alpha$ and sum of
triplet lines of He-like N VI, whereas the last column $G$ denotes
the temperature obtained from the ratio $i+f$ {\it vs} resonance
line. All the earlier results are from work of \citet{NSB02}.
Present results are slightly higher than those determined from
emissions of H- and He-like N, whereas they agree within 1$\sigma$
statistical error. The results from $G$ ratio seem to
systematically underestimate $T_{\rm e}$ as reported by
\citet{Testa04}.

Secondly, we compare present results of $n_{\rm e}$ with those
available from literatures as shown in Table 5. In available
literatures, the density of cooler plasma of stellar coronae are
usually derived from the ratio between the intercombination ($i$)
and forbidden ($f$) lines of He-like carbon, nitrogen and oxygen
ions. Above comparison has denoted the temperature of Si XI (or N
VII) emitting region is consistent with that of N VI and N VII
mixing region. So we presumed the emissions of N~VI-VII and Si~XI
are emitted by the same plasma, that is the density derived from
different techniques should be comparable. Table 5 reveals present
results agree with the previous work of \citet{NSB02} from
emissions of H- and He-like nitrogen. In the case of
$\alpha$~Cen~B, the constrained upper limit of density is
obviously lower than the density determined by \citet{NSB02}
within $1\sigma$ error, however they agree within $2\sigma$ error,
such deviation can attribute to the low S/N. In the work of
\citet{NSB02}, the effect of radiative field has been considered.

\subsection{Electron pressure and filling factor}
Large errors in ratios of Si XI result in large uncertainties in
deduced electron temperatures and densities, which imply large
uncertainties in derived electron pressures. Here, we adopt the
$n_{\rm e}$ corresponding to the best-fit value of $R_4$ to trace
the information about the pressures and structures in line forming
regions. Evaluating now the coronal pressure ($p=k_{\rm B}n_{\rm
e}T_{\rm e}$) for stars studied here as shown in Table 6. For
Procyon, present result is similar to that derived from N VI
(4.4~dyn/cm$^2$) by \citet{NMS01}. \citet{AMP03} also evaluated
the pressure using the derived $T_{\rm e}$ and $n_{\rm e}$ from
O~VII for Capella, and a upper limit of $\sim$6.0~dyn/cm$^2$ was
obtained, which is in good agreement with present result of
5.57~dyn/cm$^2$. The electron pressure of $\epsilon$~Eri appears
higher than other cases by a factor of $\sim$2, and which is also
slightly higher than that (6.60~dyn/cm$^2$) derived by
\citet{JSM01} using UV emission lines of Si~III and O~IV.

In terms of these deduced $n_{\rm e}$ and $T_{\rm e}$, we further
estimate the coronal structure through filling factors in the line
forming region. We firstly derive emitting (coronal) volumes
$V_{\rm cor}$ according to a formula $EM_{\rm ion}=0.85n_{\rm
e}^2V_{\rm cor}$, whereas $EM_{\rm ion}$ represents a ion-specific
emission measure which is determined by the correlation between
ion-specific luminosity ($L_{\rm X,ion}$) and line peak emissivity
($\epsilon_{\lambda}(T_m)$),
\begin{eqnarray}
EM_{\rm ion} & = & \frac{L_{\rm X,ion}}{\epsilon_{\lambda}(T_{\rm
m})}~~. \nonumber
\end{eqnarray}
Here, the cleanest line at 52.306~\AA\, is used to deduce the
ion-specific emission measure. For the peak emissivity, two
different values from present calculation and the Astrophysical
Plasma Emission Code (APEC) database are adopted. Secondly, we
estimate available volumes $V_{\rm avail}$ which can potentially
be filled with coronal plasma. And an assumption is made to derive
$V_{\rm avail}$, that is the plasma is confined in an uniform
distribution of loop structures obeying the loop scaling law of
\citet[hereafter RTV]{RTV78}:
\begin{eqnarray}
n_{\rm e,hot}L & = & 1.3\times10^6T^2_{\rm hot}~~\nonumber
\end{eqnarray}
with the electron density $n_{\rm e,hot}$ and the plasma
temperature $T_{\rm hot}$ at the loop top. \citet{GGS97} regarded
the loop-top temperature as equivalent to the hotter component of
a two-temperature distribution, while the temperature of the
hotter component has a such relation:
\begin{eqnarray}
T^4_{\rm hot} & =& \frac{L_{\rm
X}}{55}\left(\frac{R_{\star}}{R_{\sun}}\right)^{-2} ~~.\nonumber
\end{eqnarray}
The density at the loop-top $n_{\rm e,hot}$ we derive it from the
measured density $n_{\rm e}$(Si~XI) assuming pressure equilibrium
in a given loop, i.e., $n_{\rm e}T_{\rm ion}=n_{\rm e,hot}T_{\rm
hot}$. From these considerations, the available volumes $V_{\rm
avail}$ can be expressed as
\begin{eqnarray}
V_{\rm avail} & = &
\frac{4\pi1.3\times10^6R_{\star}^{1/2}R_{\sun}^{3/2}}{n_{\rm
e}T_{\rm ion}}\left(\frac{L_{\rm X}}{55}\right)^{3/4} \nonumber
\end{eqnarray}
in cgs units. The ratio of these volumes is defined as the filling
factor
\begin{eqnarray}
f & = & \frac{V_{\rm cor}}{V_{\rm avail}}~~.\nonumber
\end{eqnarray}

In Table 6, we derive the filling factor for Si XI emitting
regions along with the deduced electron pressures. A few percent
of the filling factor for stellar surface is derived again from
another line ratios. Present peak emissivity
(4.31$\times10^{-16}$phot.cm$^3$s$^{-1}$) is lower than the APEC
result (5.72$\times10^{-16}$phot.cm$^3$s$^{-1}$) by 25\%, which
straightforward results in the filling factor slightly higher as
indicated in Table 6. In our prediction, silicon abundance of the
solar photospheric of \citet{GGS97} is adopted, and the ionization
equilibrium calculation of \citet{MMC98} is used to deduce the
emissivity for Si~XI emission lines. The different peak emissivity
might be due to the different atomic data performed with the
different inclusion of CI. Moreover, cascade effects from higher
excited levels has been considered in present prediction.

\citet{Ness04} also derived the filling factor for a large sample
of stellar coronae using O~VII and Ne~IX lines, and very small
filling factor was found again. Assuming balance between radiation
losses and the net conductive flux, \citet{SJ03} derived {\it
area} filling factor ranging from the transition to the inner
corona for $\epsilon$~Eri. The derived temperature from Si~XI
emission is $\sim$2~MK. So the derived filling factors may be
comparable with ones from O~VII. The comparison illustrated in
Table 6 confirms this point. \citet{Testa04} also investigated
filling factors using Mg~XI and O~VII for 22 active stars. And
they concluded that the surface coronal filling factors are
$\sim10^{-4}$---$10^{-1}$ and $\sim10^{-3}$---1 for the hotter
plasma with $T_{\rm e}\sim10^7$~K and the cooler plasma with
$T_{\rm e}\sim(2-3)\times10^6$~K, respectively. Using the derived
electron temperature and density from Si~XI, filling factor of
$10^{-3}$ to $10^{-2}$ is estimated from Si~XI, which shows an
agreement with that derived from O~VII by \citet{Testa04} for
active stars.

Since the lower or upper limits of the density are obtained from
Si~XI for our sample, we further present the real constraints on
the pressure and filling factors as shown in Table 7. We estimated
the pressure in magnetic loops is larger than $\sim$0.5~dyn/cm$^2$
for our sample except for Procyon. This result is complementary to
the work of \citet{AMP03}, who derived the upper limits from O~VII
triplets. The filling factors being less than $\sim$4\% are
estimated, which shows good consistency with results constrained
by O~VII triplets. For $\alpha$~Cen~B, the lower limit of $f$
(0.36\%) and upper limit of $p$ (2.75~dyn/cm$^2$) are estimated.

\section{Conclusion}
In summary, we investigate L-shell soft X-ray emission lines of Si
XI by using a more complete kinetic model. In this model, cascade
effect from higher excited levels with $n=7$ on the upper levels
of the present interested lines has been included. Our analysis
reveals that some line intensity ratios such as $R_1$ and $R_2$,
provide useful temperature diagnostics in a temperature range of
5.5---6.5 (in log$T_{\rm e}$(K)), while $R_4$ is useful to
diagnose the density in a range of $10^{8-10}$~cm$^{-3}$.

By comparing the observed ratios with theoretical prediction, the
electron temperature and density for our sample including normal
stars (Procyon and $\alpha$~Cen~A\&B) and active binary
system---Capella, have been determined. The constrained
temperatures by $R_1$ with 1.6---2.4~MK, are consistent with the
temperature of maximum Si XI fractional abundance in the
ionization equilibrium \citep{MMC98} and the peak temperature of
the DEM constructed from a global fitting of spectra for inactive
stars. Since the blending effect in $R_2$ from Ar~IX line at
49.181~\AA\,, the correct line flux measurement of 49.207~\AA\,
may induce a certain of systemical errors. Using the experimental
ratio (1.11) of Ar~IX lines between 48.730~\AA\, and 49.181~\AA\,,
we re-derived $R_2$ and the temperatures, yet they show a good
agreement with those constrained by $R_1$. Taking into account the
contamination of Ar~IX line, densities constrained by $R_3$ also
show agreement with those by $R_4$ within $1\sigma$ statistical
error.

Temperatures determined by Si~XI emissions slightly higher than
ones constrained by emissions of H- and He-like N, whereas they
are in agreement within 1$\sigma$ error. So we presume the
emissions of N~VI-VII and Si~XI are emitted by the same plasma.
The yielded temperatures from $G$ ratio are systematically lower
than present ones as reported by \citet{Testa04}. This may be due
to the slightly lower peak temperature (1.4~MK) of the
contribution function of the N~VI triplets. An agreement of the
densities from the two different line ratios of highly charged
silicon and nitrogen, is found. This further suggests that the
Si~XI and N~VI-VII emitting plasma may be confined in the same
loop.

A filling factor of $10^{-3}$ to $10^{-2}$ is derived for our
sample from Si~XI, and a upper limit of 4\% is obtained. Such
results are consistent with the findings of \citet{Testa04} and
\citet{Ness04} that filling factors are much smaller than unity;
they also support the findings of \citet{Testa04} that the filling
factor of cooler plasma is directly proportional to the X-ray
surface flux of stars.

\begin{acknowledgements}
G. Y. thanks Dr. J.-U. Ness, Arizona State University, for his
very useful discussion. This work was supported by the National
Natural Science Foundation under Grant No. 10433010, 10403007 and
10521001.
\end{acknowledgements}

\begin{deluxetable}{llllcccccc}
\tabletypesize{\scriptsize} \tablecolumns{10} \tablecaption{Line
fluxes of interested lines for 5 late-type stars, and its
identification. The theoretical wavelength is from calculation of
FAC \citep{LZZ06c}, which is corrected by available experimental
energy levels.} \tablehead{ \colhead{} & \colhead{} &\colhead{}
&\colhead{} &\colhead{}  &
\multicolumn{5}{c}{ Line~fluxes~(in~unit~of~${\rm \times10^{-4}phot.cm^{-2}s^{-1}}$)} \\
\colhead{Index} & \colhead{$\lambda_{\rm obs}$(\AA)} &
\colhead{$\lambda_{\rm theo}$(\AA)} & \colhead{Ion} &
\colhead{Transition} & \colhead{Procyon} &
\colhead{$\alpha$~Cen~B} & \colhead{$\alpha$~Cen~A} &
\colhead{$\epsilon$~Eri} & \colhead{Capella}} \startdata 1 &
43.743& 43.757 & {\rm Si~XI} & {\rm $2s^2~^1S_0 - 2s3p~^1P_1$}
& 0.54$\pm$0.08 & 0.35$\pm$0.06 & 0.23$\pm$0.05 & 0.33$\pm$0.05 & 0.61$\pm$0.07 \\
2a & 46.283 & 46.294 & {\rm Si~XI} & {\rm $2s2p~^3P_1 -
2s3d~^3D_2$}& 0.25$\pm$0.07 & 0.19$\pm$0.05 & 0.24$\pm$0.06 & 0.15$\pm$0.04 & 0.60$\pm$0.08 \\
2b & ... & 46.260 & {\rm Si~XI} & {\rm $2s2p~^3P_0 - 2s3d~^3D_1$}   & ... & & & &  \\
2c & ... & 46.310 & {\rm Si~XI} & {\rm $2s2p~^3P_1 - 2s3d~^3D_1$}  & ... & & & &  \\
3a & 46.391 & 46.401 & {\rm Si~XI} &{\rm $2s2p~^3P_2 - 2s3d~^3D_3$} & 0.40$\pm$0.08 & 0.19$\pm$0.05 & 0.39$\pm$0.07
& 0.24$\pm$0.05 & 0.43$\pm$0.07\\
3b & ... & 46.404 & {\rm Si~XI} &{\rm $2s2p~^3P_2 - 2s3d~^3D_2$} & ... & & &  & \\
4  & 48.720 & 48.730 & {\rm Ar~IX} & {\rm $2p^53s~^1P_1 - 2p^6~^1S_0$} & 0.23$\pm$0.06 & - & 0.14$\pm$0.07 &
 0.13$\pm$0.04 & 0.16$\pm$0.05 \\
 5 & 49.207 & 49.229 & {\rm Si~XI} &{\rm $2s2p~^1P_1 - 2s3d~^1D_2$}
& 1.44$\pm$0.14 & 0.67$\pm$0.07 & 0.64$\pm$0.07 & 0.71$\pm$0.06 & 1.33$\pm$0.09 \\
6 & 52.306 & 52.295 & {\rm Si~XI} & {\rm $2s2p~^1P_1 -
2s3s~^1S_0$} & 0.75$\pm$0.11 & 0.43$\pm$0.06 & 0.39$\pm$0.07 &
0.40$\pm$0.05 & 0.73$\pm$0.08 \\
\enddata
\flushleft{{\it Notes:} The same label with different lowercase
indices (e.g., 3a and 3b) indicates blended line.}
\end{deluxetable}

\begin{table}
\centering
    \caption[I]{Diagnosed electron temperature in Si XI emitting region for 5 stars.}
    \vspace{0.2cm}
      \[
      \begin{array}{lcccc} \hline\hline
{\rm Stars} & R_1 & T_{\rm e}~{\rm (MK)} & R_2 & T_{\rm e}~{\rm (MK)} \\
\hline {\rm Procyon} & 1.39\pm0.41 & 1.65~_{-0.80}^{+1.79} &
0.64\pm0.14 &  2.24_{-1.47}^{+5.82} \\
{\rm \alpha~Cen~B} &  1.24\pm0.32 & 2.10~^{+1.80}_{-0.88} &
0.72\pm0.20 & 1.52_{-0.85}^{+5.55} \\
{\rm \alpha~Cen~A} & 1.72\pm0.57 & <2.47 & 0.76\pm0.24 &
1.23^{+4.17}_{-0.69} \\
{\rm \epsilon~Eri} & 1.20\pm0.33 & 2.30_{-1.01}^{+2.10} & 0.62\pm0.13 & 2.76_{-1.44}^{+6.53} \\
{\rm Capella} & 1.20\pm0.27 & 2.30^{+1.57}_{-0.87} &  0.67\pm0.12
& 2.02_{-0.94}^{+3.03} \\ \hline
         \end{array}
      \]
      \end{table}
\begin{table}
\centering
    \caption[I]{Diagnosed electron density (in unit of $\times10^9$~cm$^{-3}$) in Si XI emitting region for our sample.}
    \vspace{0.2cm}
      \[
      \begin{array}[h]{lcccc} \hline\hline
{\rm Stars} & R_3 & n_{\rm e}~ & R_4 & n_{\rm e} \\
\hline {\rm Procyon} & 2.95\pm0.86 & <87.86 & 1.88\pm0.65 &  >0.03 \\
{\rm \alpha~Cen~B} &  3.16\pm0.88 & <5.00 & 2.26\pm0.32 & <0.95 \\
{\rm \alpha~Cen~A} & 2.13\pm0.82 & >0.52 & 1.62\pm0.52 & >0.44 \\
{\rm \epsilon~Eri} & 2.46\pm0.73 & >0.25 & 1.65\pm0.55 & >0.42 \\
{\rm Capella} & 2.78\pm0.67 & >0.06 &  1.71\pm0.51 & >0.33 \\
\hline
         \end{array}
      \]
      \end{table}
\begin{table}
\centering
    \caption[I]{Present results of temperature and those derived from H- and He-like N ions \citep{NSB02}.
     Ly$\alpha$/He represents the flux ratio between the Ly$\alpha$ line and the sum of triplet lines of N~VI.}
    \vspace{0.2cm}
      \[
      \begin{array}[h]{lcccc} \hline\hline
     & \multicolumn{4}{c}{T_e~{\rm (MK)}} \\
{\rm Stars} &{\rm Present} &  {\rm Ly}\alpha/r & {\rm Ly}\alpha/{\rm He} & G=\frac{i+f}{r} \\
\hline {\rm Procyon} & 1.65~_{-0.80}^{+1.79} & 1.09\pm0.13 & 1.41\pm0.07 & 1.28\pm0.46 \\
{\rm \alpha~Cen~B} & 2.10~^{+1.80}_{-0.88}   & 1.79\pm0.16 &
1.43\pm0.22 & <1.8 \\
{\rm \alpha~Cen~A} & <2.47  & 1.35\pm0.09 & 1.15\pm0.09
& <1.2 \\
{\rm \epsilon~Eri} & 2.30_{-1.01}^{+2.10}  & 2.47\pm0.18 & 1.95\pm0.30 & <2.4 \\
{\rm Capella} & 2.30^{+1.57}_{-0.87} & 2.73\pm0.08 & 2.02\pm0.10 & 0.46\pm0.28 \\
\hline
         \end{array}
      \]
      \end{table}
\begin{table}
\centering
    \caption[I]{Present results of electron density and those derived from
     the ratio between $i$ and $f$ lines of He-like N VI ion \citep{NSB02}.}
    \vspace{0.2cm}
      \[
      \begin{array}[h]{lcc} \hline\hline
     & \multicolumn{2}{c}{{\rm log}~n_{\rm e}/{\rm cm^{-3}}} \\
{\rm Stars} &{\rm Present} &  R=\frac{i+f}{r} \\
\hline{\rm Procyon} & >7.48  & 9.96\pm0.23 \\
{\rm \alpha~Cen~B} & <8.98 & 9.99\pm0.65 \\
{\rm \alpha~Cen~A}  & >8.64 & 9.95\pm0.30 \\
{\rm \epsilon~Eri}  & >8.62 & 10.35\pm0.33\\
{\rm Capella} & >8.52 & 9.86\pm0.12 \\
\hline
         \end{array}
      \]
      \end{table}
\begin{table}
\centering
    \caption[I]{Filling factor and electron pressures $p$ derived from ${\rm n_e}$
     corresponding to the best-fit value of $R_4$.
    For comparison, the filling factor from O~VII are also listed \citep{Ness04}.}
    \vspace{0.2cm}
      \[
      \begin{array}[h]{lcccc} \hline\hline
     & p & \multicolumn{3}{c}{{\rm filling~factor~{\it f}~in~\%}} \\
{\rm Stars} &{\rm dyn/cm^2} & {\rm Pres.} & {\rm APEC} &  {\rm O~VII} \\
\hline{\rm Procyon} & 2.69 & 3.78 & 2.84 & 1.31 \\
{\rm \alpha~Cen~B} & 0.70 & 1.44 & 1.09 & 0.62 \\
{\rm \alpha~Cen~A}  & 5.71 & 0.14 & 0.10 & 0.18 \\
{\rm \epsilon~Eri}  & 10.86 & 0.25 & 0.19 & 0.62 \\
{\rm Capella} & 5.57 & 0.45 & 0.34 & 0.64 \\
\hline
         \end{array}
      \]
      \end{table}
\begin{table}
\centering
    \caption[I]{Real constraints on the pressure $p$ (in dyn/cm$^2$) and the filling factor $f$ (in \%).}
    \vspace{0.2cm}
      \[
      \begin{array}[h]{l|ccccc} \hline\hline
 &{\rm Procyon} & {\rm \alpha~Cen~B} & {\rm \alpha~Cen~A} &  {\rm \epsilon~Eri} & {\rm Capella}
 \\\hline
p & >0.07 & <2.75 & >0.57 & >1.33 & >1.05 \\
f& - & >0.36 & <1.40 & <2.04 & <3.30 \\
\hline
         \end{array}
      \]
      \end{table}

\clearpage
\begin{figure}
\centering
\includegraphics[height=8.8cm,angle=-90]{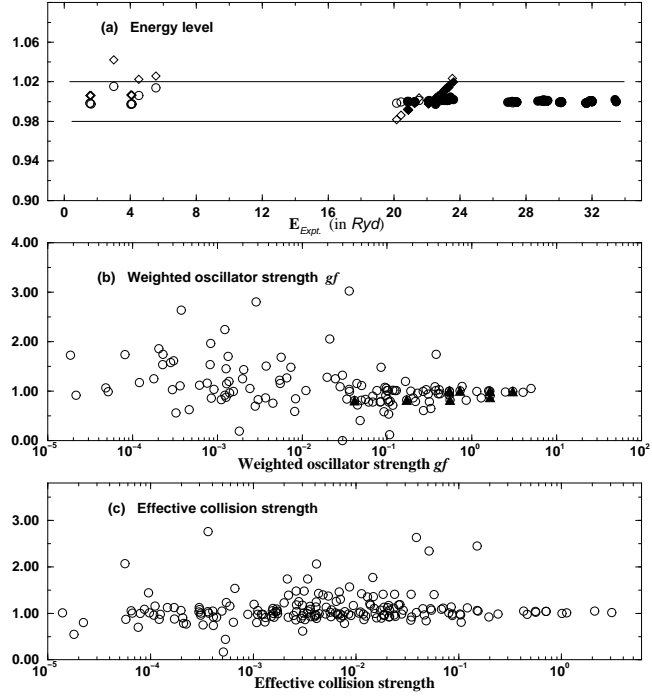}
\caption[short title]{Comparison of our calculated atomic data
with other available data. (a) Comparison of different theoretical
energy levels with available experimental ones (in unit of {\it
Ryd}). x-axis is the experimental energy from NIST (open symbols)
or the results of \citet{CT01} (filled symbols), while y-axis
represents ratios of different calculation {\it vs} experimental
one. Different symbols correspond to the ratio of results of
different calculations {\it vs} experimental ones. $\circ$:
present results, $\diamond$: theoretical data in the CHIANTI
database. The solid horizontal lines represents 2\% uncertainty
range. (b, c) Comparison of our results with the input data of
CHIANTI code. x-axis is the present result, while y-axis denotes
the ratio of the input data of the CHIANTI code {\it vs} present
one. Filled up-triangle denotes the transitions interested in this
work.}
\end{figure}
\begin{figure}
\centering
\includegraphics[angle=-90,width=8cm]{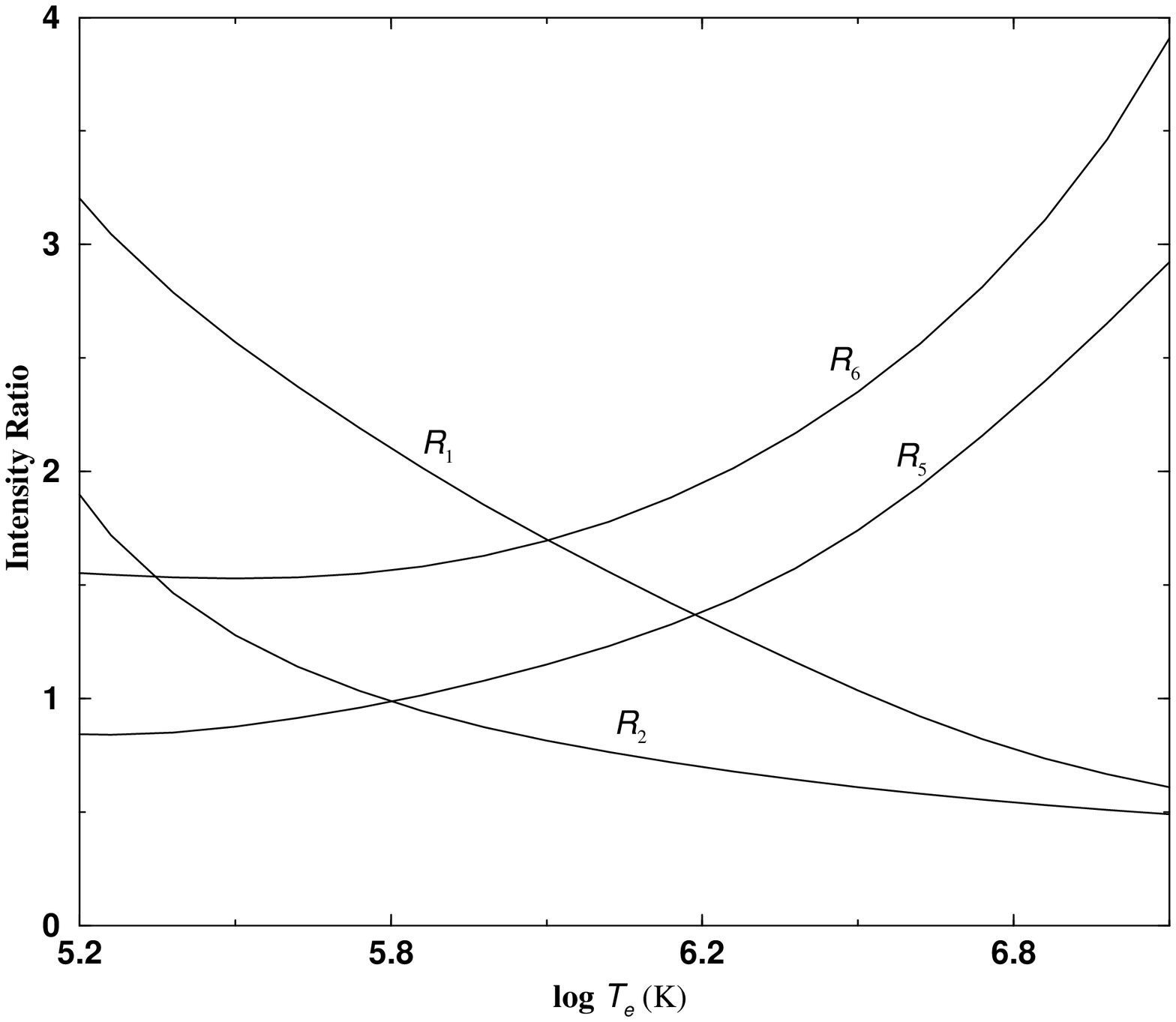} \\
\includegraphics[angle=-90,width=8cm]{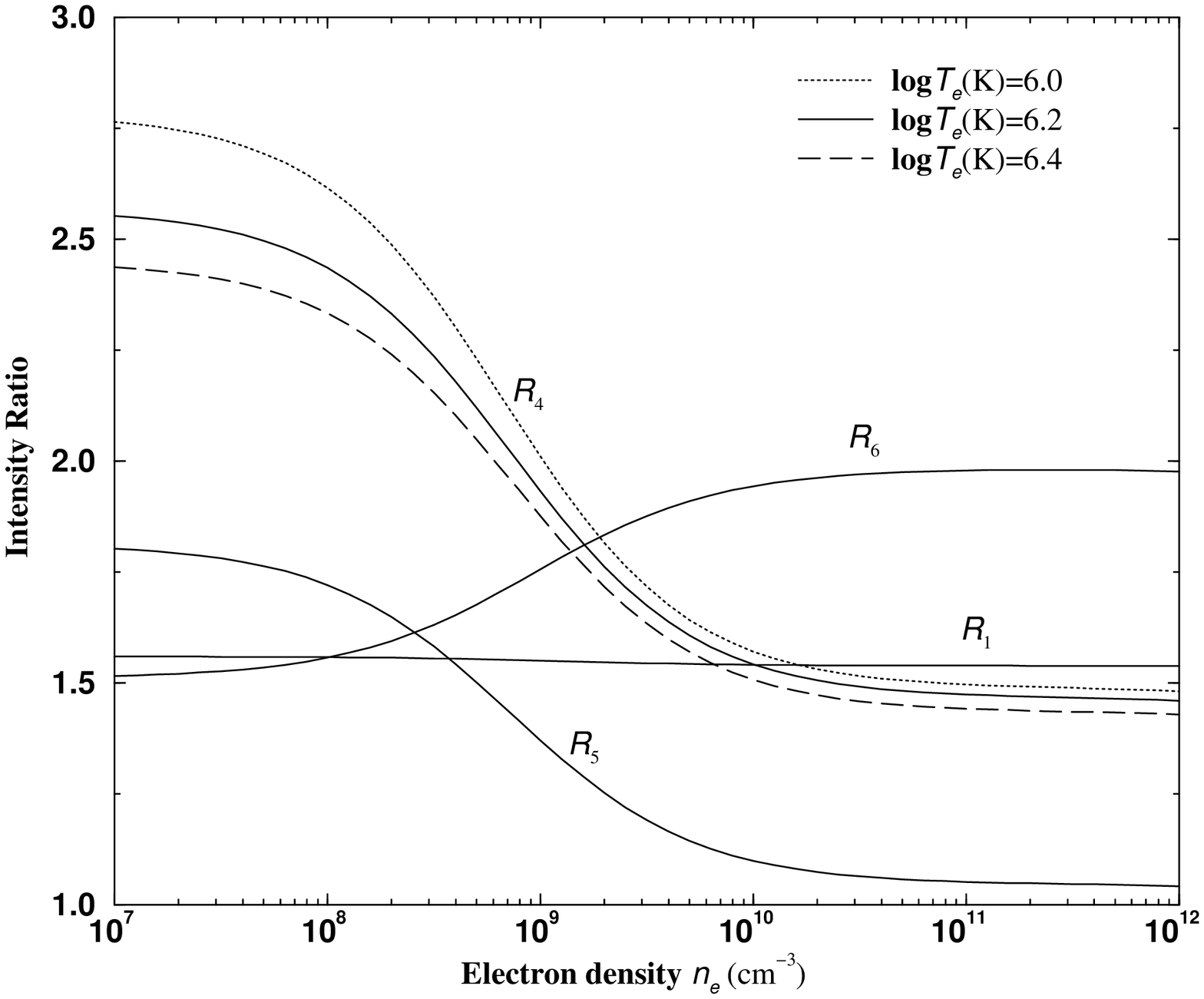}
\caption[short title]{The theoretical line ratios of Si XI. {\it
Upper}: plotted as a function of the electron temperature at
$n_{\rm e}$=1.0$\times$10$^9$~cm$^{-3}$. {\it Lower}: plotted as a
function of the electron density, solid line corresponding to
log$T_{\rm e}$(K)=6.2.}
\end{figure}
\begin{figure}
\centering
\includegraphics[angle=-90,width=8cm]{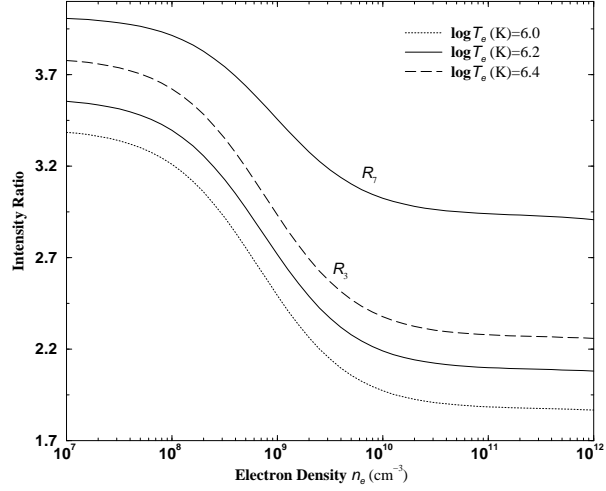}
\caption[short title]{The theoretical Si XI emission line ratios
{\it vs} the electron density. The solid line corresponding to
log$T_{\rm e}$(K)=6.2.}
\end{figure}
\begin{figure*}
\centering
\includegraphics[angle=0,width=8cm,clip]{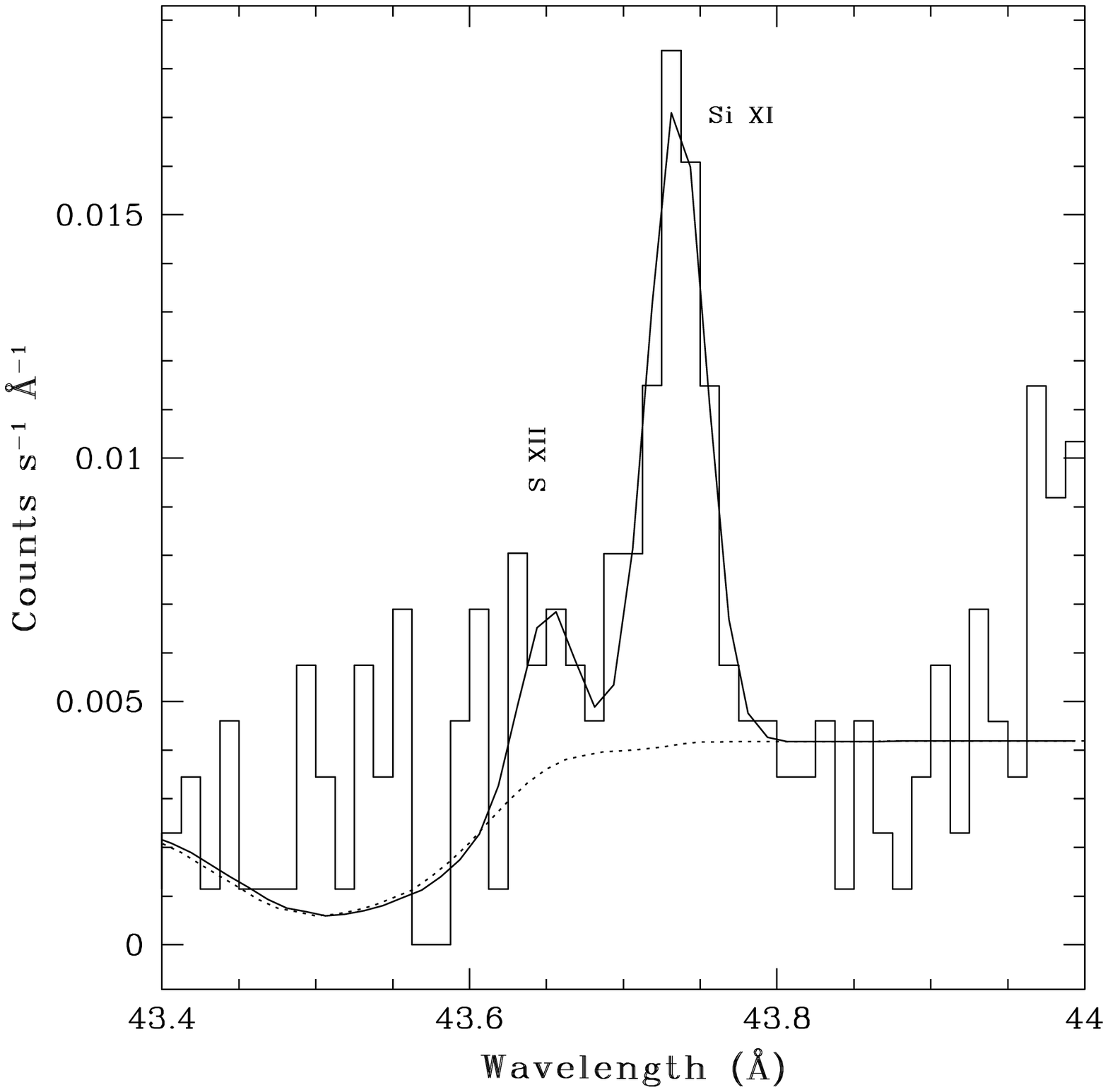} %
\includegraphics[angle=0,width=8cm,clip]{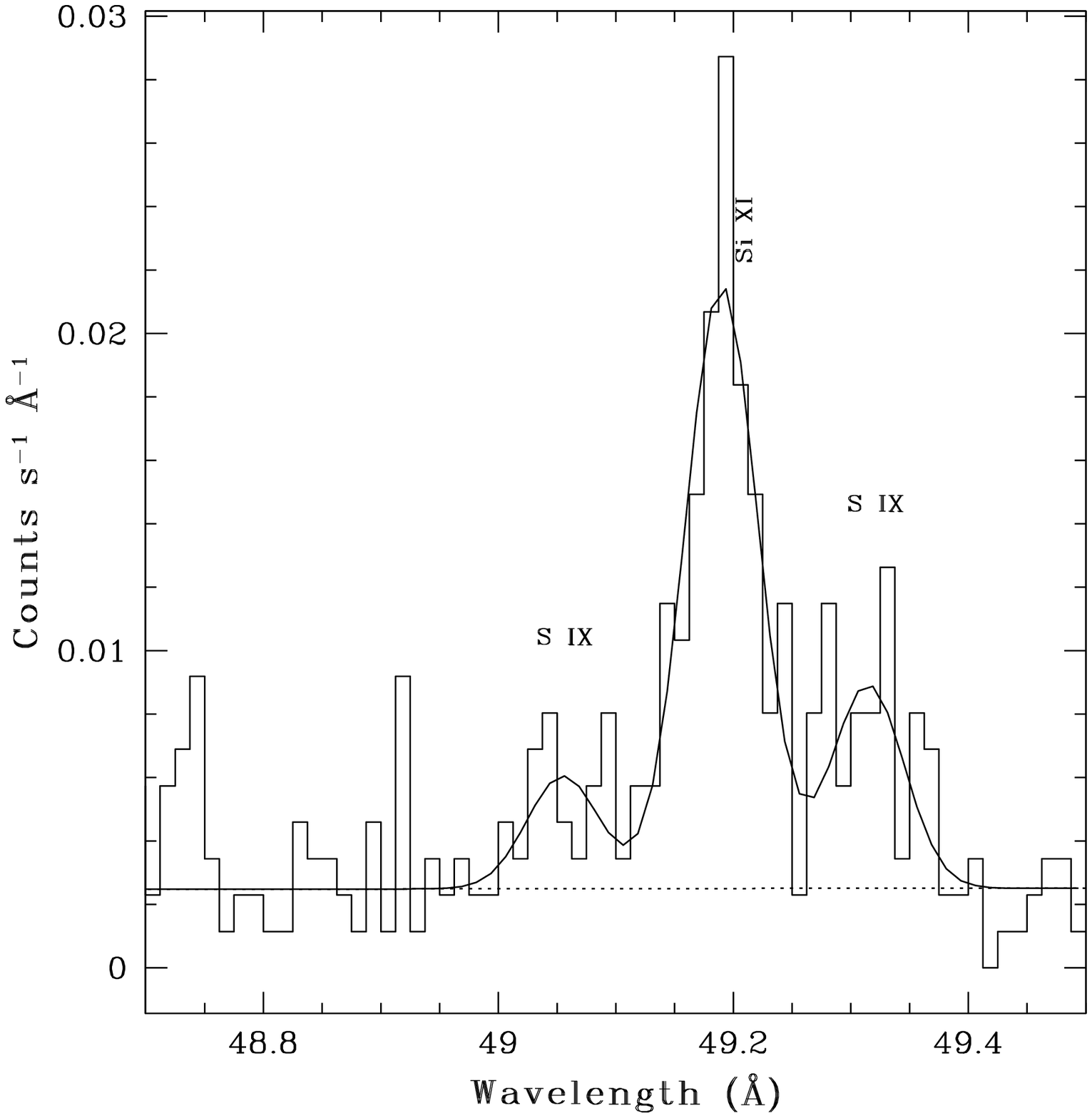}
\caption[short title]{Spectrum of Procyon observed with LETGS and
best-fit spectra (smooth lines) in wavelength ranges of
43.4---44.0~\AA\, and 48.7---49.5~\AA\,. Dot lines represent the
continuum levels. Prominent lines have been labelled above curves.
}
\end{figure*}
\begin{figure}
\centering
\includegraphics[angle=-90,width=8cm]{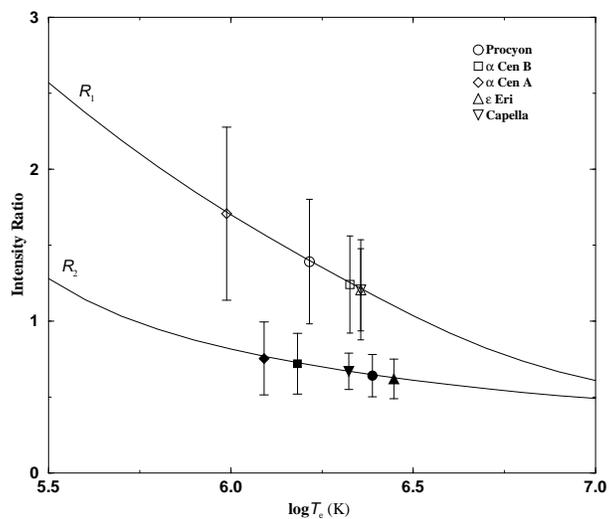}
\caption[short title]{Comparison between the theoretical line
intensity ratios ($R_1$ and $R_2$) and observed ratios of five
stars. Opened symbols with error bars are observed values of
$R_1$, while filled symbols are values of $R_2$. }
\end{figure}
\begin{figure}
\centering
\includegraphics[angle=-90,width=8cm]{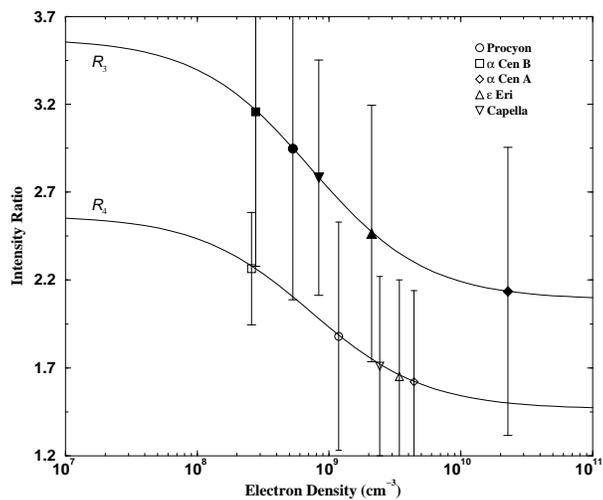}
\caption[short title]{Comparison between the theoretical line
intensity ratios ($R_3$ and $R_4$) and observed ratios of our
sample. Opened symbols with error bars represent observed values
of $R_3$, while filled symbols are for $R_4$. }
\end{figure}
\end{document}